# Collection of remote optical signals by air waveguides

E. W. Rosenthal, N. Jhajj, J. K. Wahlstrand, and H. M. Milchberg[*]

Institute for Research in Electronics and Applied Physics, University of Maryland, College Park, MD 20742
*Corresponding author: milch@umd.edu

**Collection of weak signals from remote locations is the primary goal and the primary hurdle of optical stand-off detection schemes. Typically, the measured signal is enhanced using large numerical aperture collection optics and high gain detectors. We show that the signal in remote detection techniques can be enhanced by using a long-lived air waveguide generated by an array of femtosecond filaments. We present a proof of principle experiment using an air plasma spark source and a ~1 m air waveguide showing an increase in collected signal of ~50%. For standoff distances of 100 m, this implies that the signal-to-noise ratio can be increased by a factor ~$10^4$.**

In optical stand-off detection techniques, spectroscopic or other light-based quantitative information is collected from a distance. Among the most popular schemes are light detection and ranging (LIDAR) and laser-induced breakdown spectroscopy (LIBS). In LIDAR, the signal is induced by a laser pulse, either by reflection or backscattering from distant surfaces or atmospheric constituents. In remote LIBS, laser breakdown of a distant target is accompanied by isotropic emission of characteristic atomic and ionic species. Some recent schemes for optical stand-off detection use femtosecond filamentation, which occurs when an ultrashort pulse, propagating through a transparent medium such as the atmosphere, experiences focusing from its self-induced Kerr lens [1]. When self-focusing is stronger than beam diffraction, the beam mode collapses into a tight core or filament where the local intensity reaches the ionization threshold of the medium. The beam collapse is then arrested by plasma defocusing. The interplay of focusing and defocusing over the pulse temporal envelope leads to extended propagation of the filament core over distances typically exceeding many Rayleigh lengths. The ability to deliver high peak intensities at relatively long distances has been applied to LIDAR [2] and LIBS [3], and other remote sensing schemes [4,5]. Other applications of filaments include THz generation [6,7], supercontinuum and few-cycle pulse generation [8,9], channeling of electrical discharges [10], and microwave guiding [11].

Recently, we have studied the long time hydrodynamic response of the gas through which a filament has propagated [12, 13]. Unique to femtosecond filaments is their extended high intensity propagation over many Rayleigh lengths and their ultrafast nonlinear absorption in the gas, stored in plasma and atomic and molecular excitation [12-16]. This creates an axially extended impulsive pressure source to drive gas hydrodynamics. After the filamenting pulse passes, a gas density depression or hole grows over several hundred nanoseconds. Over the same timescale, a single cycle acoustic wave is launched and begins to propagate away from the hole, which then slowly decays by thermal diffusion over milliseconds. Understanding this evolution has led to our use of symmetric filament arrays to generate very long-lived air waveguide structures [14] that can support very high power secondary laser pulses. On microsecond timescales, the colliding acoustic waves launched by the filament array form a fiber-like guiding structure with a gas density (or refractive index) enhancement in the center [14, 16]. On timescales well past the acoustic response, the residual gas density holes thermally relax and spread over milliseconds, forming the cladding of a long lived fiber-like structure with higher gas density in the center. Even a single filament's acoustic wave [17] can trap light in an annular mode over a microsecond time window, as we have recently explained [16].

In this paper we show that femtosecond filament-generated air waveguides can collect and transport remotely-generated optical signals while preserving the source spectral shape. The air waveguide acts as an efficient standoff lens. Here, we demonstrate collection of an isotropically emitted optical signal, the worst case scenario in terms of collection efficiency. Even stronger collection enhancement would apply to directional signals from stimulated backscattering [18] or backward lasing [5]. Our results have immediate impact on remote target applications of laser-induced breakdown spectroscopy (LIBS) [3] and on LIDAR studies [2]. Our proof of principle experiment tests ~1 m long air waveguides of various configurations, in both the acoustic and thermal regimes. Simple extrapolation of our results to >100 m waveguides generated by extended filamentation [1,19,20] implies potential signal-to-noise enhancements greater than ~$10^4$.

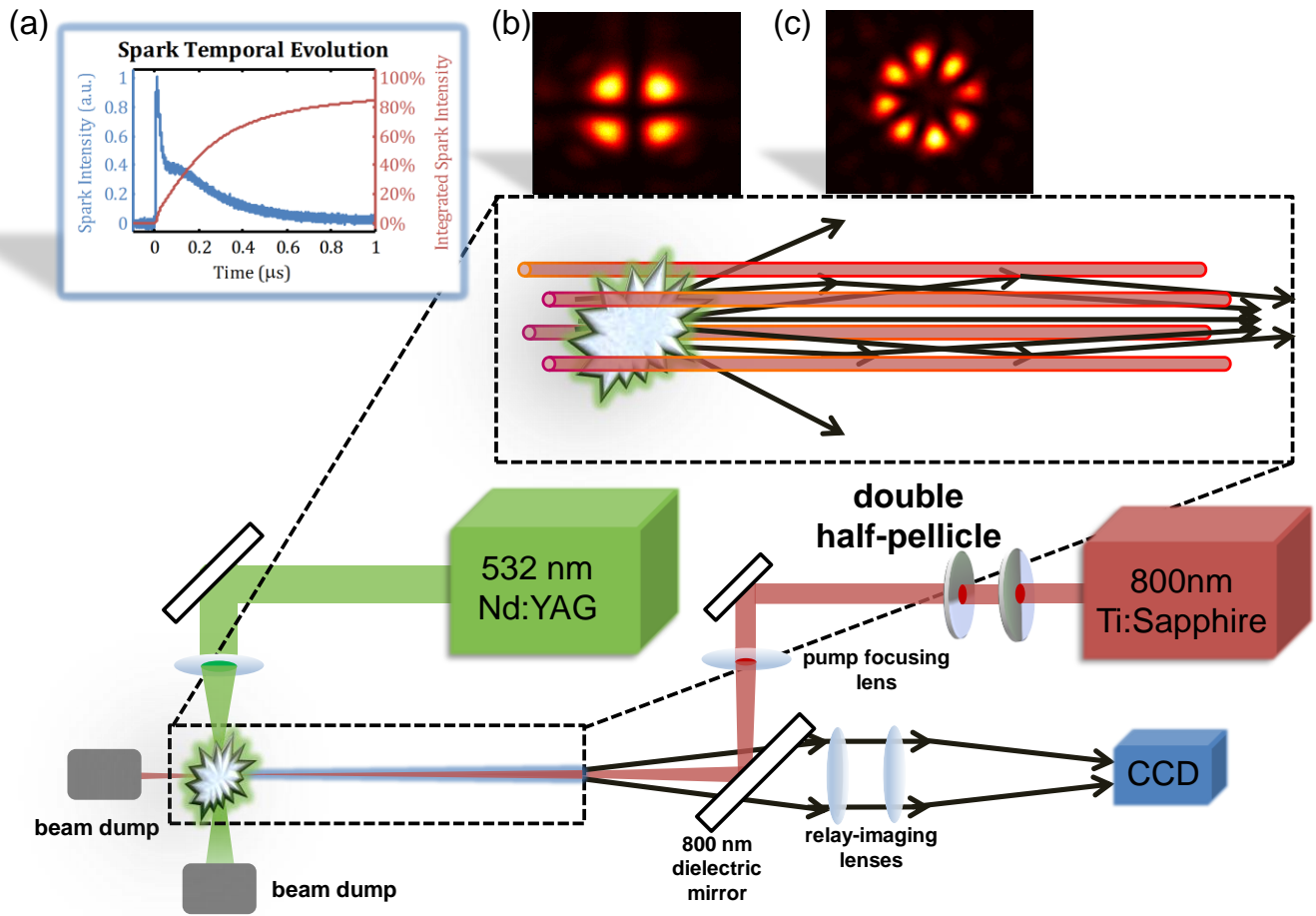

Fig. 1. Experimental setup for demonstration of light collection and transport by the air waveguide. Insets: (a) Time evolution of the light emitted by the spark. (b) Low intensity image of the 4 lobe beam focus generated by orthogonal half pellicles. (c) Low intensity image of the 8 lobe beam focus generated by the segmented mirror.

In remote LIBS, as a specific example, laser-breakdown of a gas or solid target of interest generates a characteristic line spectrum that allows identification of target constituents. However, as the optical emission from the target is isotropic with a geometrical $R^{-2}$ falloff with source distance, very little of the signal is collected by a distant detector, necessitating large numerical aperture collection optics and high gain detectors [3]. Schemes to increase the LIBS signal by increasing the plasma temperature and/or density have been proposed, such as use of double pulses [21], but all such methods are still subject to the geometrical factor.

Figure 1 illustrates the experimental setup. Single filaments and filament arrays 75-100 cm long are generated in air using 10 Hz Ti:Sapphire laser pulses at 800 nm, 50-100 fs, and up to 16 mJ. The beam focusing is varied between $f/400$ and $f/200$ depending on the type of guide. Arrays with four or eight filaments are generated by phase shifting alternating segments of the beam's near field phase front by $\pi$. As described in our earlier work, four-filament arrays, or quad-filaments, are generated using two orthogonal "half-pellicles" [14], and 8-filament arrays, or octo-filaments, are generated using eight segment stepped mirrors [16], resulting in either a $TEM_{11}$-like mode or a linear combination of $LG_{0,\pm 4}$ modes in the low intensity beam focus, as seen in insets (b) and (c) of Fig. 1. Above the self-focusing threshold, the beam lobes collapse into parallel and distinct filaments whose cores maintain the relative phase relationship imposed by the pellicles or the stepped mirror [14, 16]. By inserting burn paper into the paths of the quad- and octo-filaments [14], we have verified that their 4 or 8 lobe character is preserved through the whole propagation path. As described above, colliding acoustic waves at the array center launched by either quad-filaments or octo-filaments form waveguides of duration $\sim 1$ μs, roughly corresponding to the acoustic wave transit time through the array center. Millisecond lifetime waveguides develop during the slow post-acoustic thermal diffusion of the density holes left by the filaments [14].

We test the signal collection properties of our waveguides using an isotropic, wide bandwidth optical source containing both continuum and spectral line emission, provided by tight focusing at $f/10$ of a 6 ns, 532 nm, 100 mJ laser pulse to generate a breakdown spark in air. Time evolution of the spectrally integrated spark emission is shown in an inset of Fig. 1. The signal FWHM is $\sim 35$ ns, with a long $\sim 1$ μs decay containing >85% of the emitted light energy. The air spark laser and the filament laser are synchronized with RMS jitter <10 ns. The delay between the spark and the filament structure is varied to probe the time-evolving collection efficiency of the air waveguides. The air spark and filament beams cross at an angle of 22°, so that the spark has a projected length of $\sim 500$ μm transverse to the air waveguide. As depicted in Fig. 1, the spark is positioned just inside the far end of the air waveguide. Rays from the source are lensed by the guide and an exit plane beyond the end of the guide is imaged through an 800 nm dielectric mirror onto a CCD camera or the entrance slit of a spectrometer. This exit plane is located within 10 cm of the end of the waveguide.

The collected signal appears on the CCD image as a guided spot with a diameter characteristic of the air waveguide diameter. Guided spots are shown in Fig. 2 for five types of air waveguide: the quad-filament and octo-filament waveguides in both the acoustic and thermal regimes, and the single filament annular acoustic guide. Surrounding the guided spots are shadows corresponding to the locations of the gas density depressions, which act as defocusing elements to scatter away source rays. We quantify the air waveguide's signal collecting ability using two measures. The peak signal enhancement, $\eta_1$, is defined as the peak imaged intensity with the air waveguide divided by the light intensity without it. We define the source collection enhancement, $\eta_2$, as the integrated intensity over the guided spot, divided by the corresponding amount of light on the same CCD pixels in the absence of the air waveguide. Figure 3 shows plots of both $\eta_1$ and $\eta_2$ for each of our waveguide types as a function of time delay between the spark and filament laser pulses. Inspection of inset (a) in Fig. 1 shows that ~70% of the spark emission occurs before 500 ns, so the evolution of the peak signal and collection enhancements are largely characteristic of the waveguide evolution and not the source evolution. The spot images shown in Fig. 2 are for time delays where the collection efficiency is maximized for each waveguide. In general, we find $\eta_1 > \eta_2$ because the peak intensity enhancement is more spatially localized than the spot.

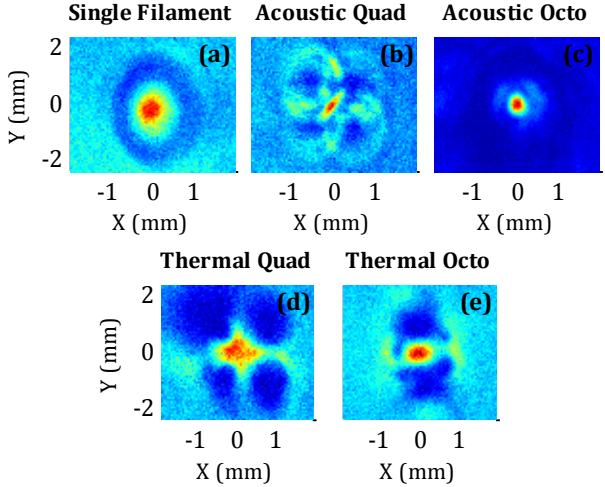

Fig. 2. Single shot images of the breakdown spark light emerging from the exit of the guiding structures. (a) Single filament -induced guide at 1.2 μs, (b) Four-lobed acoustic guide at 3.2 μs, (c) Eight-lobed acoustic guide at 1.4 μs, (d) Four-lobed thermal guide at 250 μs, (e) Eight-lobed thermal guide at 100 μs,

As can be seen from Figs. 2 and 3, and as discussed earlier, there are two temporal regimes for guiding. In the case of filament arrays, a guide is formed when the acoustic waves from individual filaments collide on axis, creating an on-axis index increase surrounded by the index decrement from the density holes corresponding to the quad- and octo-filaments (Fig. 2(b) and (c)). This takes place on a ~1 μs timescale, as borne out by the associated plots of evolution of signal enhancement in Fig. 3(b) and (c). After several hundred microseconds, the thermal waveguide develops. By that point, the diffusion of heat deposited by the filament array has formed a moat of hotter, lower density gas acting as a cladding surrounding a central core of relatively unperturbed gas. The associated plots of peak and collection enhancement for both the quad-filament and octo-filament thermal guides show an almost 2 ms long collection window, ~10³ times longer than for the acoustic guides. For a single filament, we also see source light trapping in a window ~1 μs long, where trapping occurs in the positive crest of the single cycle annular acoustic wave launched in the wake of the filament [16]. Here, the trapping lifetime is constrained by the limited temporal window for source ray acceptance as the acoustic wave propagates outward from the filament.

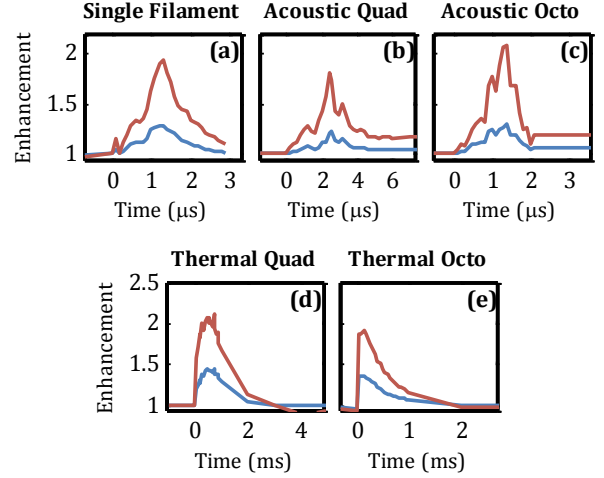

Fig. 3. Source collection enhancement (blue) and peak signal enhancement (red) plotted vs. filament - spark source delay for (a) single filament acoustic guide, (b) four-lobed acoustic guide, (c) eight-lobed acoustic guide, (d) four-lobed thermal guide, and (e) eight-lobed thermal guide.

For each of the guides we observe peak signal enhancement in the range $\eta_1$ ~1.8–2, and source collection enhancement $\eta_2$ ~1.3–1.5. (Fig. 2(a-e) and Fig. 3 (a-e)). For a waveguide numerical aperture of $NA = \sqrt{2}(\delta n_{co} - \delta n_{cl})^{1/2}$ [14], where $\delta n_{co}$ and $\delta n_{cl}$ are the shifts in the air waveguide core and cladding refractive indices relative to undisturbed ambient air, the source collection enhancement is

$$\eta_2 = 4(NA)^2 \left(\frac{a_{in}}{w}\right)^2 \left(\frac{L}{a_{out}}\right)^2 \alpha \qquad (1)$$

where $L$ is the waveguide length, $a_{in}$ and $a_{out}$ are the mode diameters for the waveguide at the spark source and the output respectively, $w$ is the greater of the source diameter $d_s$ and $a_{in}$, and $\alpha$ is a transient loss coefficient computed from a beam propagation method [22] simulation. For our octo-thermal guide we used burn paper to characterize the transverse profile of the guide, and a microphone [16] to measure the axial extent. Similar to previous experiments [14], we find the mode diameter to be roughly half the lobe spacing. Parameters for the octo-thermal guide are $NA$~2.5·10⁻³, $L$ = 1 m, $a_{in}$ = 0.5 mm, $a_{out}$ = 1.5 mm, $d_s$ = 0.5 mm, and $\alpha$ = 0.28, resulting in $\eta_2$ = 3, which is in reasonable agreement with our measured $\eta_2$ = 1.4. Although the signal enhancement is modest for our meter-scale filament, the scaling $\eta_2 \propto L^2$ would enable a ~10⁴ collection enhancement by a 100 m air waveguide.

For the BPM simulation, we used a paraxial portion of a spherical wave to simulate rays from a point source lensed by an eight-lobed thermal index structure with peak index shift consistent with the NA used above. Substantial transient losses are observed over the first ~50cm which are factored into Eq. (1) through $\alpha$. Between ~1 and 20 m of propagation through the guide, high order leaky modes decay away, leaving dominant, very slowly leaking modes to propagate over the remaining distance. Attenuation coefficients for these dominant modes are < 0.4 dB/km. After transient losses over the first meter, 25% of the signal is lost over the remaining 99 m of transit. These results do not account for absorption in air.

Crucial to remote sensing schemes is identification of source chemical composition, which is typically done by identifying characteristic spectral lines of neutral atoms or ions of a given species. For such schemes, it is important that the emission spectrum at the source location be conveyed with high fidelity to the remote detector location. To investigate this property of our air waveguide, we compared spectra measured 10 cm from the air spark source to spectra of the guided signal collected from the output of the air waveguide. The results are shown in Fig. 4, for a thermal waveguide from a quad-filament, where the spectra have been integrated over 100 shots. There is no significant difference in the spectra except for attenuation in the range 700-900 nm owing to signal transmission through a broadband 800 nm dielectric mirror, as depicted in Fig. 1, and the onset of UV absorption at less than ~330nm, possibly due to the accumulating ozone generated by the 10 Hz spark and filament waveguide sources. In addition to characteristic nitrogen emission lines identified in Fig. 4, a very strong scattering peak is seen at 532 nm from the spark laser, with the spectral peak extending well past the range of the plot's vertical axis.

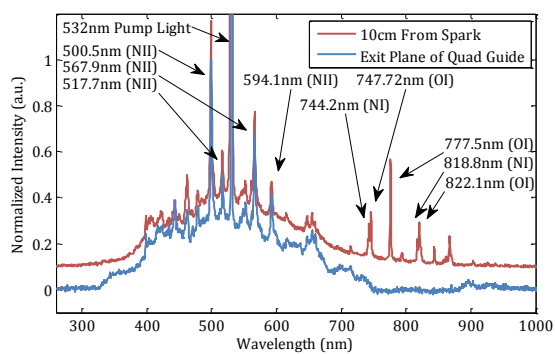

Fig. 4. Air-spark spectrum collected near the source (red curve) and after transport in a 75 cm air waveguide (thermal guide from a quad-filament, blue curve). Characteristic lines are indicated on the spectrum. The red curve is raised for clarity. (Spectrometer: Ocean Optics HR2000+)

In conclusion, we have demonstrated that a femtosecond filament-generated air waveguide can be used as a remote broadband collection optic to enhance the signal in standoff measurements of remote source emission. This provides a new tool for dramatically improving the sensitivity of optical remote sensing schemes. By employing air waveguides of sufficient length, the signal-to-noise ratio in LIDAR and remote LIBS measurements can be increased by many orders of magnitude. Finally, we emphasize that air waveguides are dual purpose: not only can they collect and transport remote optical signals, but they can also guide high peak and average power laser drivers to excite those sources.

## Acknowledgments
This research was supported by the Air Force Office of Scientific Research, the Defense Threat Reduction Agency, and the National Science Foundation. The authors thank R. Birnbaum and S. Zahedpour for useful discussions and technical assistance.